\documentclass[twocolumn,pre,aps,showpacs,amsmath,amssymb]{revtex4-1}
\usepackage{graphicx}
\usepackage{color}
\usepackage{bm}
\usepackage{soul}
\usepackage{ulem}
\graphicspath{{NewFigures/}}

\begin{document}
	
	\title{Controlling fluid adhesion force with electric fields}
	\author{Pedro H. A. Anjos$^1$}
	\email{pamorimanjos@iit.edu}
	\author{Francisco M. Rocha$^2$}
	\email{francisco.rocha@univ-amu.fr}
	\author{Eduardo O. Dias$^3$}%
	\email{eduardo.dias@ufpe.br}
	\affiliation{%
		$^1$Department of Applied Mathematics, Illinois Institute of Technology, Chicago, Illinois 60616, USA}
	\affiliation{%
		$^2$Aix Marseille University, CNRS, Institut Universitaire des Systèmes Thermiques et Industriels, 13454 Marseille, France}
	\affiliation{%
		$^3$Departamento de F\'{\i}sica, Universidade Federal de Pernambuco,
		Recife, PE  50670-901, Brazil}

	
	\begin{abstract}
	Developing adhesives whose bond strength can be externally manipulated is a topic of considerable interest for practical and scientific purposes. In this work, we propose a method of controlling the adhesion force of a regular fluid, such as water and/or glycerol, confined between two parallel plates by applying an external electric field. Our results show the possibility of enhancing or diminishing the bond strength of the liquid sample by appropriately tuning the intensity and direction of the electric current generated by the applied electric field. Furthermore, we verify that, for a given direction of the electric current, the adhesion force can be reduced enough for the fluid to lose its adhesive properties and begin exerting a force to move apart the confining plates. In these circumstances, we obtain an analytical expression for the minimum electric current required to detach the plates without requiring the action of an external force.
	\end{abstract}
	\maketitle
	
	\section{Introduction}
	\label{intro}

	Many areas of science and technology, including chemistry, rheology, and hydrodynamics, are applied to the understanding and development of adhesive materials~\cite{Gay1,Aubrey}. Even though, strictly speaking, fluids are not \lq\lq genuine\rq\rq~adhesives, they have similar physical features and behaviors to regular soft adhesive materials. Therefore, although fluid properties are generally much simpler than the rheological characteristics of conventional adhesives, the use of fluids helps to gain insight into the adhesion features of these more complex materials. For these reasons, the study of the bond strength of fluids has been a theme of plenty of recent research~\cite{Francis,Anke,Gay,Dias,Mir7,Mir8,anjosforce,Russel,Gay2,Mir6,Ben,Anke2,Anke3,Van,Barral,Randy,Diasnew3,new1,new2,new3,new4,new5,new6}.
	
	The existing theoretical and experimental studies of fluid adhesion forces include Newtonian, non-Newtonian (shear-thinning, shear-thickening,
	viscoelastic, etc.), and magnetic fluids (ferrofluids and magnetorheological fluids). The measure of the adhesive strength of spatially constrained, liquid thin films is provided by the so-called probe-tack test~\cite{Zosel,Lakrout}. A typical version of this procedure takes place with a fluid sample confined between two closely spaced parallel flat plates. While the upper plate is lifted at a constant velocity, the applied force is measured. The behavior of this force as a function of the upper plate displacement is used to quantify the adhesive strength of the liquid sample.
	
	An essential aspect of common interest, regardless of the type of fluid involved, is the pursuit of controlling its bond strength. In this scenario, the protocol proposed in Ref.~\cite{Dias} refers to the adhesion energy of Newtonian and power-law fluids required to separate the two plates of the probe-tack test. More specifically, given that one wishes to lift the apparatus’ upper plate until a certain height in a fixed time interval, Ref.~\cite{Dias} obtains the optimal time-dependent lifting drive function for which the adhesion energy is minimized. 
	
	Another controlling mechanism, which is more related to the approach presented in this work, is proposed in Refs.~\cite{Mir7,Mir8}. They provide an efficient and reversible way to either enhance or reduce the adhesion force of magnetic liquids by varying the intensity and the geometric configuration of an externally applied
	magnetic field. In Ref.~\cite{Mir7}, the authors use ferrofluids, which are colloidal suspensions of nanometer-sized magnetic particles in a nonmagnetic carrier fluid. They show that for perpendicular (out-of-plane) or azimuthal (in-plane) magnetic fields, the adhesive force of the ferrofluid is always reduced. On the other hand, if the applied magnetic field lies on the plane of the plates but points radially outwards, the adhesive strength can be either enhanced or reduced.
	
	In Ref.~\cite{Mir8}, another type of magnetic liquid known as magnetorheological (MR) fluid is utilized as an adhesive sample. MR fluids typically consist of micron-sized magnetized particles dispersed in aqueous or organic carrier liquids. The results presented in Ref.~\cite{Mir8} show that the bond strength of these fluids is hugely increased when a uniform perpendicular (out-of-plane) or a nonuniform radial magnetic field is utilized. Furthermore, it is verified that this enhancement is due to the field-dependent nature of the yield stress.
	
	In this work, we propose a method of controlling fluid adhesive properties by exploiting an electro-osmotic flow~\cite{Bazant1, Bazant2, Bazant3, pedroeletro,mengeletro} generated in the probe-tack test by an applied electric field. In contrast to the magnetic control of Refs.~\cite{Mir7,Mir8}, our strategy does not require any particular type of fluids, such as ferrofluids and MR fluids. As a matter of fact, it can be implemented by using regular fluids, e.g., water and/or glycerol.

	
	Electro-osmotic flow arises over electrically charged surfaces due to the interaction of an externally applied electric field with the net charge in the electric double layer (EDL)~\cite{EDL1,EDL2,EDL3}. More specifically, when the liquid sample is brought into contact with the probe-tack apparatus' glass plates, their surfaces become negatively charged due to the dissociation of ionic surface groups. Consequently, a thin, diffuse cloud of excess counter-ions in the liquid accumulates near the surfaces, forming the so-called EDL. By applying an external electric field, these mobile ions are subjected to a net electric force, which drives an electro-osmotic flow, in addition to the flow driven by the lifting process. Since the intensity and direction of the external electric field are easily manipulated, the induced secondary electro-osmotic flow can either assist or oppose the flow driven by the lifting process. As we will verify in the following sections, these features enable the development of adhesives whose bond strength is regulated by electric means.
	
	\section{Derivation of fluid adhesion force under the influence of electric fields} 
	\label{derivation}
	
	\begin{figure}
		\includegraphics[width=0.49\textwidth]{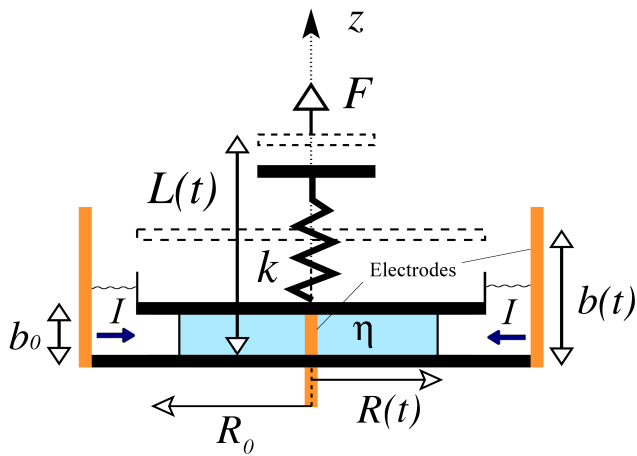}
		\caption{Representative sketch of the probe-tack apparatus, where a Newtonian inner fluid of viscosity $\eta$ is confined between parallel plates. A surrounding fluid with negligible viscosity is added to conduct current to the inner fluid. A radial electric current $I$, whose intensity and direction can be manipulated, is applied by circular electrodes positioned at the center and the outer edge of the apparatus.}
		\label{geom}
	\end{figure}
	
	Figure~\ref{geom} sketches the probe-tack apparatus subjected to a radial electric field. An incompressible, Newtonian fluid with viscosity  $\eta$, permittivity $\varepsilon$, and (surface) potential $\zeta$ is confined between two parallel glass plates of the apparatus. Our goal is to control the adhesion force of this fluid by applying an electric field that generates an electro-osmotic flow, supplementing the flow driven by the lifting process. To this end, we introduce circular electrodes positioned at the center and the outer edge of the plates. In addition, we add a surrounding fluid in contact with the external electrode to conduct the electric current to the fluid of interest (see Fig.~\ref{geom}). The outer fluid is appropriately chosen to have a much smaller viscosity than $\eta$; hence, as it will be clearer later, its contribution to the adhesion force can be neglected. For this reason, from now on, our discussion will be focused on the inner fluid.

	The initial plate-plate distance is represented by $b_{0}$ and the initial fluid radius by $R_0$. While the lower plate is held fixed at $z=0$, with the $z$-axis pointing in the upward direction perpendicular to the plates, an external force $F$ is applied at the upper plate. Following Refs.~\cite{Francis,Anke,Gay,Mir7}, we assume that the apparatus has a spring constant denoted by $k$.  The plate spacing evolves in time according to $b=b(t)$, with the deformation due to the stretching of the apparatus being $L(t) - b(t)$. Thus, the control parameter that is set by the motors of the flexible apparatus in real-world probe-tack experiments is $L=L(t)$, and not $b$. For the rigid apparatus case (situation not considered here), we have $L=b$. Here at $t=0$, we have that $b_{0}=L_{0}$. Lastly, an external, in-plane radial electric field ${\textbf E}=-\bm{\nabla} \Phi$ (with $\Phi$ being the electric potential within the fluid) is applied, yielding a radial electric current $I=I(t)$ parallel to the flow direction (see Fig.~\ref{geom}).

	The goal of this section is to calculate the pulling force $F$ as a function of the displacement $L$, taking into account both hydrodynamic and electric contributions. Following Ref.~\cite{Anke} and many subsequent works, we derive $F$ considering that the interface between the fluids remains circular during the lifting process, with a time-dependent radius $R=R(t)$. This approach is justified by the fact that experiments of Ref.~\cite{Anke} with highly developed fingering structures are very well described by theoretical force-distance curves which assume a circular interface. Nevertheless, we direct the interested readers to Ref.~\cite{anjosforce}	for situations where viscous fingering development is relevant to the adhesion problem. The conservation of fluid volume leads to 
	\begin{equation}
		\label{cons}
		\pi(R^2-R_e^2)b=\pi(R^2_0 -R_e^2)b_0,
	\end{equation}
	where $R_e$ is the radius of the inner electrode.

	Since most experimental and theoretical studies in probe-tack adhesion~\cite{Francis,Anke,Russel,Gay,Gay2,Mir6,Mir7,Ben,Anke2,Anke3,Van,Mir8,Barral,Randy} deal with very small $b$, relatively low lifting velocities, and highly viscous fluids, effects of fluid inertia can be safely neglected. For more details about the circumstances in which the fluid inertia is relevant in probe-tack adhesion, we refer the reader to Ref.~\cite{Diasnew3}. In addition, as we focus on a high aspect ratio $R/b \gg 1 $, a situation traditionally considered (see, e.g., Ref.~\cite{Ben}) in probe-tack tests, we adopt a Darcy-law-like approach, which involves the dynamic of the gap-averaged velocity $\textbf v$.
	
	As mentioned earlier, the contact of the fluids with the glass plates of the probe-tack apparatus dissociates ionic surface groups, which causes the glass surface to become negatively charged. As a result, a thin, diffuse cloud of excess counter-ions in the liquids accumulates near the surface, forming the EDL~\cite{EDL1,EDL2,EDL3}. Therefore, besides the flow driven by the lifting process, the external electric field ${\textbf E}$ yields a net electric force on the mobile ions, driving an electro-osmotic flow. This electric-induced flow can be either in the same (inward) or opposite (outward) direction of the lifting-driven  flow (hydraulic flow), depending on the direction of ${\textbf E}$.

	Under the circumstances above, for the quasi-two-dimensional  plate-plate  geometry, the motion of the fluids, with contribution of electric forces, is described by the modified Darcy's law~\cite{EDL2,Bazant1,Bazant2,Bazant3} for the $z$-averaged velocity
	\begin{equation}
		\label{Darcy1}
		{\textbf v} =-M\bm{\nabla}p -K\bm{\nabla}\Phi,
	\end{equation}
	where $M=b^{2}/(12 \eta)$ and $K=-\varepsilon\zeta/\eta$ (Helmholtz-Smoluchowski relation~\cite{EDL2}) are the hydraulic and electro-osmotic mobilities, respectively,  and $p$ denotes the pressure field. The first and second terms on the right-hand side (RHS) of Eq.~(\ref{Darcy1}) represent, respectively, the pressure-driven and the electro-osmotic contributions to the flow. In addition, the total current density ${\textbf J}$ is the sum of streaming (pressure-induced) and Ohmic (driven by the electric field) currents 
	\begin{equation}
		\label{Darcy2}
		{\textbf J} =-K\bm{\nabla}p -\sigma\bm{\nabla}\Phi,
	\end{equation}
	where $\sigma$ is the fluid Ohmic conductivity. To conclude all governing equations of our system, we have the $z$-average of the three-dimensional incompressibility condition,
	\begin{equation}\label{incomp}
		{\bm \nabla}\cdot {\bf v}=-\frac{\dot b}{b}.
	\end{equation}

	Since there is neither injection nor withdrawal, the fluid velocity ${\bf v}$ must vanishes at the inner electrode, $r=R_e$. Then, as we have assumed an circular interface, i.e., the flow is symmetric in the azimuthal direction, the solution of Eq.~(\ref{incomp}) is
	\begin{equation}\label{incompsol}
		v_r=-\frac{\dot b}{2b}\left(r-\frac{R_e^2}{r}\right).
	\end{equation}
	Notice that this expression can also be obtained by taking the time derivative of the volume conservation~(\ref{cons}) and setting $v_r={\dot R}$.
	
	For the situation where the upper plate is lifted at a constant velocity, the external force ${\bf F}=F{\bf e_z}$ should balance the fluid pressure force ${\bf F}_{p}=F_{p}{\bf e_z}$, where ${\bf e_z}$ is the upward unit vector. Thus, $F$ can be calculated by integrating the pressure field $p$ over the region occupied by the fluid, i.e.,
	\begin{equation}\label{force}
		F=-F_{p}=-\int_{R_e}^R(p-p_0)2\pi r dr,
	\end{equation}
	where $p_0$ is the pressure of the outer fluid, which is approximately constant since its viscosity is much smaller than $\eta$. Hence, $p_0$ is approximately equal to the atmospheric pressure. At this point, it is worth mentioning that, for conventional lifting processes without electro-osmotic flow, the pressure difference $p-p_0$ is negative, hence $F_p < 0$, and the fluid acts as a sort of glue between the parallel plates. As a result, in order to keep the lifting velocity constant, $F$ must be positive~(a lifting force), being, therefore, a measure of the fluid's adhesive strength. Nevertheless, we will see that, $p - p_0$ (and also $F_p$) may become positive due to the electro-osmotic flow, meaning that the fluid ceases to be an adhesive and instead acts to separate the plates. In this case, $F < 0$ measures the separation force performed by the liquid.
	
	Since we need the pressure field to compute the applied force~(\ref{force}), we combine Eqs.~(\ref{Darcy1}) and~(\ref{Darcy2}) in such a way to eliminate the dependence on ${\bm \nabla} \Phi$, so that 
	\begin{equation}\label{gradient}
		{\bm \nabla} p=-\frac{1}{\sigma M-K^2}\left(\sigma {\bf v} -K{\bf J}\right).
	\end{equation}
	From this equation, we verify that the fluid velocity $v_r$, which was obtained in Eq.~(\ref{incompsol}), and the current density $J_r$ are necessary to calculate the pressure gradient. Furthermore, note that by substituting $M=b^2/(12\eta)$ and $K=-\varepsilon\zeta/\eta$ into Eq.~(\ref{gradient}), and making the fluid viscosity small enough, we have ${\bm \nabla} p\approx 0$. This is what we are assuming for the outer fluid. 
	
	\begin{table*}[ht]
		\setlength{\tabcolsep}{10pt}
		\begin{tabular}{@{}c c c c c c c c@{} }
			\hline
			\hline
			$\eta$ & $\varepsilon_0$ & $\varepsilon$ & $\zeta$ & $\sigma$ & $R_0$ & $R_e$ & $k$ \\
			$109$ mPa\,s & $8.8 \times 10^{-12}$ F/m & $49.1\varepsilon_0$ & $-150$ mV & $155 \times 10^{-4}$ S/m & $2.0$ cm & $1.0$ mm & $1.0 \times 10^{-5}$ N/m\\
			\hline
			\hline
		\end{tabular}
		\caption{Material parameters used throughout the paper for the confined fluid sample, as well as geometrical and elasticity parameters of the probe-tack apparatus. The material properties of the fluid utilized here, including electrical characteristics, are consistent with the experimental works in Refs.~\cite{Anke,Ben,Anke4,Leider,Bazant2}. Additionally, the probe-tack parameters are based on existing adhesion tests with viscous fluids~\cite{Francis,Anke,Russel,Gay,Gay2}.}
		\label{table:Params}
	\end{table*}

	To compute $\textbf J$, first we combine Eqs.~(\ref{Darcy1}) and~(\ref{Darcy2}) again, but now we eliminate the dependence on ${\bm \nabla} p$, and take the divergence operator on both sides of the resulting expression. Then, using ${\bm \nabla}^2 \Phi=0$, we obtain
	\begin{equation}\label{JeV}
		{\bm \nabla}\cdot{\textbf J}=\frac{K}{M}{\bm \nabla}\cdot{\textbf v}.
	\end{equation}
	Substituting the two-dimensional incompressibility condition~(\ref{incomp}) into Eq.~(\ref{JeV}), the divergence of $\textbf J$ can be written as
	\begin{equation}\label{current1}
		{\bm \nabla}\cdot{\textbf J}=-\frac{K}{M}\frac{\dot b}{b}.
	\end{equation}
	 Notice that the three-dimensional effect of the upper plate lifting on the velocity $\bf v$, illustrated on the RHS of Eq.~(\ref{incomp}), is transferred to $\bf J$, as shown on the RHS of Eq.~(\ref{current1}). It is worth mentioning that in previous studies concerning electro-osmotic flows (see, for instance, Refs.~\cite{Bazant1,Bazant2,pedroeletro}), ${\bm \nabla}\cdot{\textbf J}=0$ as the plates are held fixed ($\dot b = 0$). Using the fact that the current density at the inner electrode ($r=R_e$) is given by $J_r(r=R_e)=I/(2\pi b R_e)$, the solution of Eq.~(\ref{current1}) is 
	\begin{equation}\label{current}
		J_r=-\frac{K}{M}\frac{\dot b}{2b}\left(r-\frac{R_e^2}{r}\right)+\frac{I}{2\pi b r}.
	\end{equation}
	From the relation between $J_r(r=R_e)$ and $I$ above,  $I>0$ ($I<0$) means that the current is directed outward (inward), i.e., opposite (same) to the hydraulic flow. Substituting Eqs.~(\ref{incompsol}) and~(\ref{current}) into Eq.~(\ref{gradient}), and integrating the resulting expression over $r$, we obtain
	\begin{eqnarray}\label{pressure}
		p&=&-\frac{\dot b}{4bM}\big(R^2-r^2\big)\nonumber\\
		&+&\left(\frac{K }{\sigma M-K^2}\frac{I}{2\pi b}+\frac{{\dot b}R_e^2}{2bM}\right)\ln{\left(\frac{r}{R}\right)} + p_0,
	\end{eqnarray}
	where we have used the boundary condition $p(R)=p_0$. Finally, we can compute the force $F$ under the influence of an electric field by substituting Eq.~(\ref{pressure}) into Eq.~(\ref{force}). Using $M=b^2/(12\eta)$, $K=-\varepsilon\zeta/\eta$ (where $\zeta <0$), and the conservation of fluid volume shown in Eq.~(\ref{cons}), $F$ can be written as
	\begin{eqnarray}\label{forcef}
		F=\frac{3 \eta b_0^2 (A_0-A_e)^2 {\dot b}}{2\pi b^5}+\frac{3\eta}{\pi b}\left(\frac{\varepsilon |\zeta| I}{\eta \sigma b^2 - 12 \varepsilon^2 \zeta^2}-\frac{A_e {\dot b}}{b^2}\right)\nonumber\\
		\times\left[(A_0-A_e)\frac{b_0}{b}- A_e\ln\left(1+\frac{b_0(A_0-A_e)}{A_e b}\right)\right],\nonumber\\
	\end{eqnarray}
	%
	where $A_0=\pi R_0^2$ and $A_e=\pi R_e^2$ are the initial area of the fluid and the cross-sectional area of the inner electrode, respectively. In Eq.~(\ref{forcef}), let us call the first (second) term the hydraulic (electro-osmotic) contribution to the applied force $F$. Note that the time dependence of $F$ in Eq.~(\ref{forcef}) is implicit in $b$, $\dot b$, and can also be in $I$ if one considers time-dependent currents. Taking the limit where the radius of the electrode is small, i.e., $R_e \rightarrow 0$, Eq.~(\ref{forcef}) simplifies significantly, becoming 
	\begin{eqnarray}\label{forcef1}
		F&=&\frac{3 \eta b_0 A_0}{2\pi b^2}\left(\frac{b_0 A_0 \dot{b}}{b^3} + \frac{2 \varepsilon |\zeta| I }{\eta \sigma b^2 - 12 \varepsilon^2 \zeta^2} \right).
	\end{eqnarray}
	%
	Furthermore, the traditional adhesion force of Newtonian fluids~\cite{Anke} without the influence of an electric field is recovered if one considers $I=0$ in Eq.~(\ref{forcef1}).
	
	As mentioned before, due to the compliance of the measurement apparatus, the actual plate spacing $b$ is not necessarily equivalent to $L$. The presence of the compliance yields an interplay between viscous, electrohydrodynamic force, as quantified by $F=F(b, {\dot b})$, and the spring restoring force $k(L - b)$, which results from the deflection of the apparatus. Neglecting the apparatus inertia, the acceleration drops out from this dynamic scenario. As a result, the dynamic is determined by the nonlinear, first-order ordinary differential equation (ODE) for $b(t)$~\cite{Francis,Gay,Anke,Russel,Gay2}
	\begin{eqnarray}
		\label{completeF} 
		k (L - b)=F(b, {\dot b}).
	\end{eqnarray}

	 In the following section, using the full or simplified expression for the applied force [Eqs.~\eqref{forcef} or~\eqref{forcef1}, respectively], and the force balance~\eqref{completeF}, we will compute $F$ as a function of $L$ for a given set of physical and geometrical parameters. Then, we will investigate how the coupling between the hydrodynamic and electro-osmotic flows allows us to control the fluid adhesive strength.
	
	\section{Results} 
	\label{disc}
	
	We proceed to investigate how the presence of the electric field, and hence the electric current, modifies the viscous adhesion force. The parameters used hereafter in this paper, which are consistent with existing probe-tack setups and experimental works on electro-osmotic fluid flow~\cite{Francis,Gay,Anke,Russel,Gay2,Mir6,Mir7,Ben,Anke2,Anke3,Van,Mir8,Barral,Randy,Diasnew3,Bazant2}, are shown in Table~\ref{table:Params}. Moreover, as conventionally done in adhesion tests with viscous fluids~\cite{Francis,Gay,Anke,Russel,Gay2,Mir6,Mir7,Ben,Anke2,Anke3,Van,Mir8,Barral,Randy,Diasnew3}, all the analysis performed here considers that the upper plate moves with constant speed $V$, such that $L(t) = b_0 + Vt$.
	
	\begin{figure}[!t]
		\includegraphics[width=0.48\textwidth]{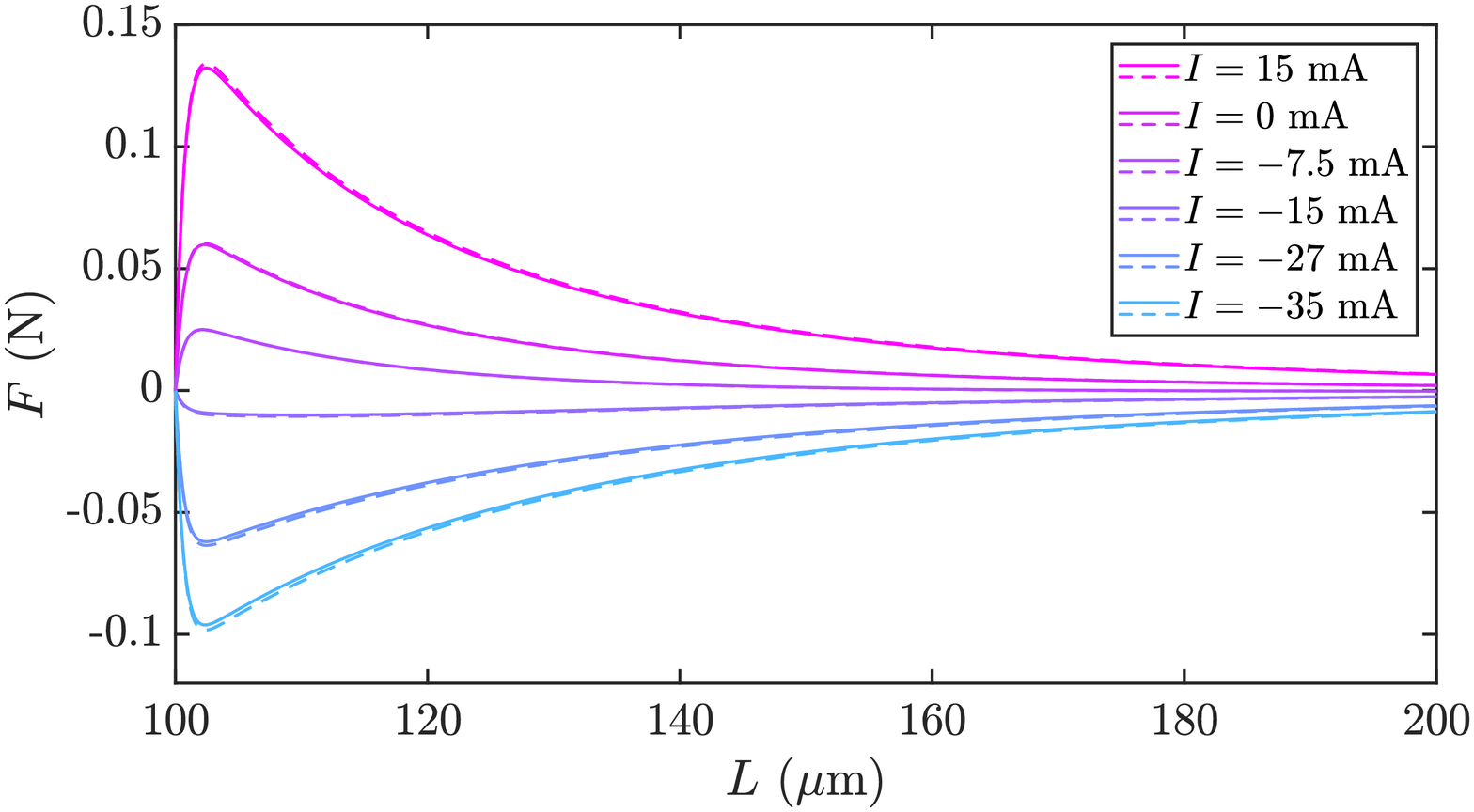}
		\caption{Applied force $F$ as a function of the displacement $L = b_0 + V t$, for different values of the applied (constant) electric current $I$. Solid lines represent the adhesion force obtained via Eq.~\eqref{forcef}, while dashed curves are computed with~\eqref{forcef1}, where $R_e \rightarrow 0$. For all cases the lifting velocity and initial gap are kept fixed and equal to $V = 0.8$ $\mu$m/s and $b_0 = 100$ $\mu$m, respectively.}
		\label{fig:FxL}
	\end{figure}
	
	\begin{figure*}
		\includegraphics[width=\textwidth]{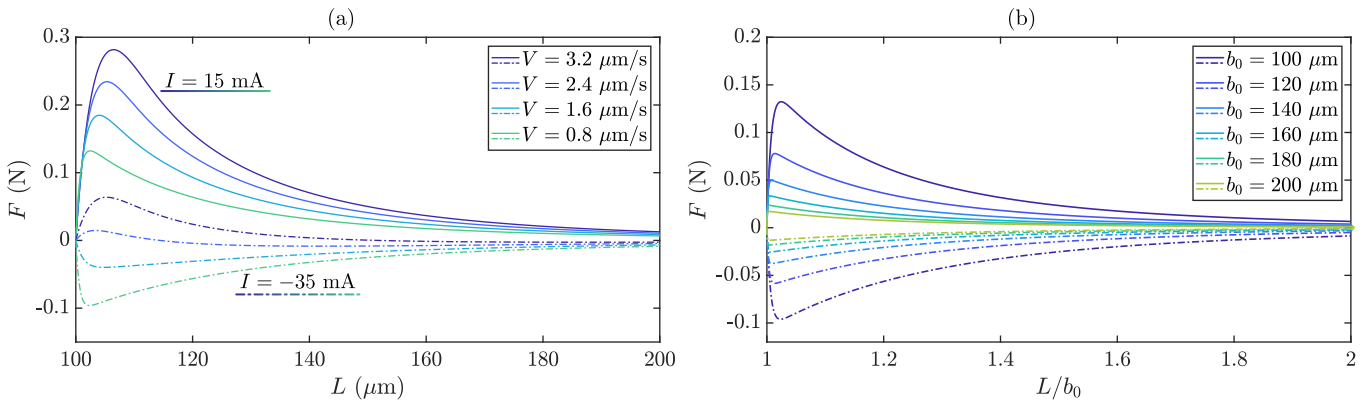}
		\caption{The applied force $F$ as a function of the displacement $L$ for two different electric currents, $I = 15$ mA (solid lines) and $I = -35$ mA~(dashed lines). In panel (a), $F$ is plotted as a function of the displacement $L$ for different values of $V$ and a fixed initial gap $b_0 = 100$ $\mu$m. On the other hand, in panel (b), $F$ is plotted against $L/b_0$ for different initial gaps $b_0$ and a fixed lifting velocity $V = 0.80$ $\mu$m/s.}
		\label{fig:FmaxDiffVsB0s}
	\end{figure*}

	In order to obtain $F$ as a function of $L$, we first substitute the expression~\eqref{forcef} [or its simplified version, Eq.~\eqref{forcef1} for $R_e \rightarrow 0$] into the force balance~\eqref{completeF}. The resulting ODE for the time evolution of the gap $b=b(t)$ is solved numerically, and then, we substitute this numerical outcome appropriately back into~\eqref{forcef} or~\eqref{forcef1}. Thus, we finally obtain the force $F$ as a function of $L$, as well as in terms of the initial gap $b_0$, the lifting velocity $V$, the electric current $I$, and the material and setup parameters given in Table~\ref{table:Params}. ~
	
	We start investigating the behavior of $F$ as a function of the displacement $L$, for a fixed initial gap $b_0 = 100$ $\mu$m, and lifting velocity $V = 0.8$ $\mu$m/s. The results are depicted in Fig.~\ref{fig:FxL} for six constant values of electric current. Solid lines represent the calculations using~\eqref{forcef}, whilst dashed lines are obtained by considering Eq.~\eqref{forcef1} ($R_e \rightarrow 0$ limit). As in the traditional situation of probe-tack test with viscous fluids, i.e., $I = 0$, when the elasticity of the apparatus is taken into account, the adhesion force for $I \neq 0$ does not peak at $b = b_0$~\citep{Francis}. Instead, at $t = 0$, $L = b$ and the force is zero. Soon after the test starts, the gap between the plates, $b$, remains roughly constant, and the motion is dominated by changes in $L$, which leads to an approximately linear increase in $F$ [see Eq.~\eqref{completeF}]. Then, during the lifting, force is transmitted to the gap $b$, and after a while, its rate of change $\dot b$ equates to the lifting velocity $V$ so that the force reaches its peak value. This peak is often called the adhesive strength of the material in the case of positive peaks. At long times, the spring deformation is quite small so that $b \approx L$ and $\dot{b} \approx V$, and thus $F \sim L^{-5}$.

	By inspecting Fig.~\ref{fig:FxL}, we verify that for $I > 0$, hydraulic and electro-osmotic contributions add up, resulting in an overall increase of the adhesion force $F$ ($F>0$), as one can observe by comparing the curves for $I=0$ and $I=15$ mA. The physical reason for this enhancement in the adhesion strength is similar to the one given in Ref.~\cite{Bazant2} to explain the increase in the pressure gradient due to an electro-osmotic effect on injection-driven flows. When $I>0$, the applied electric field acts on the positive ions in the EDL, driving an electro-osmotic flow in the opposite direction to the (inward) hydraulic flow. Thus, the pressure gradient must increase to maintain the flow rate imposed by the volume conservation~(\ref{cons}) during the lifting process, compensating for the opposite electro-osmotic flow. This compensation causes an apparent viscosity enhancement, which Ref.~\cite{Bazant2} refers to as apparent ``electrokinetic thickening.''

	The increase of the pressure gradient can be observed by analysing Eq.~(\ref{gradient}) and noticing that $\sigma M > K^2$ and $J_r>0$ when $I>0$ for the physical parameters of Table I and electric currents we used. This augmented pressure gradient reduces the fluid pressure and thus enhances the difference $p_0-p$ in Eq.~(\ref{force}), resulting in a larger applied force $F$. Lastly, in the expression of $F$ [Eqs.~\eqref{forcef} and~\eqref{forcef1}], one can easily verify this behavior by noticing that $\eta \sigma b^2 \gg 12 \epsilon^2 \zeta^2$. Hence, similar to the hydraulic contribution, $I > 0$ yields a positive influence of the electro-osmotic term on $F$.

	On the other hand, when $I<0$, the term proportional to the electric current in Eqs.~\eqref{forcef} and~\eqref{forcef1} becomes negative, creating a competition between the hydraulic and electric-induced contributions. The consequence of this competition is weakening the required force to pull the plates apart~(e.g., $I=-7.5$ mA in Fig.~\ref{fig:FxL}). A physical explanation for this situation is analogous to the one for $I>0$, but now, with the electro-osmotic flow being in the same direction as the hydraulic flow, resulting in an apparent ``electrokinetic thinning.''

	If one increases the magnitude of $I$, with $I<0$, such that the electro-osmotic term is dominant in Eq.~(\ref{force}), the applied force becomes negative~(e.g., $I = -27$ mA and $-35$ mA in Fig.~\ref{fig:FxL}). In this situation, the flow of the mobile ions increases the fluid pressure enough to create a net force that acts to move the plates apart. As a consequence, in order to keep the lifting velocity constant, one needs to exert a downward (resistive) force~($F<0$) on the upper plate of the probe-tack apparatus. Also, it should be pointed out that the plates would detach accelerating upwards if no resistive force were applied.

	\begin{figure*}
		\includegraphics[width=\textwidth]{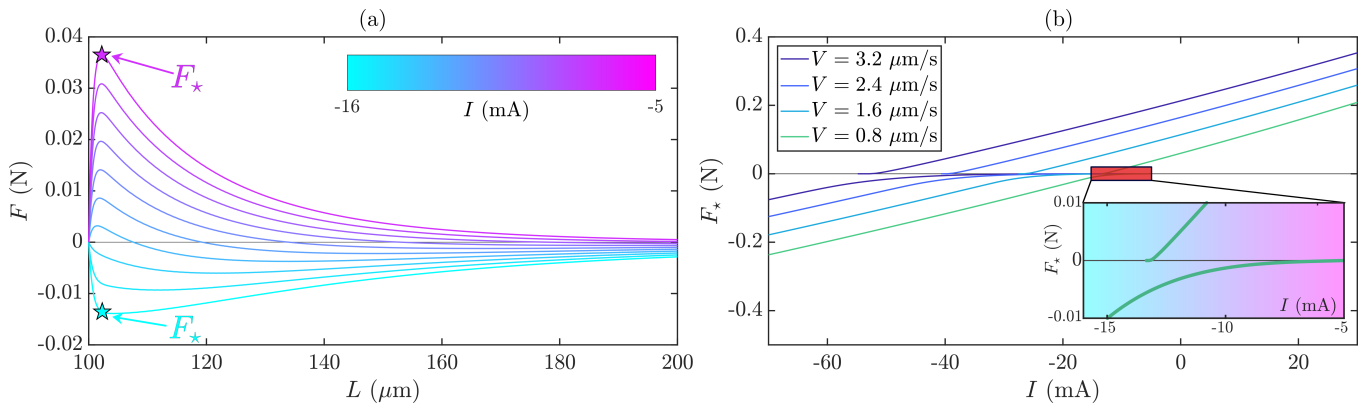}
		\caption{(a) The applied force $F$ as a function of the displacement $L$ for $I \in [-16,-5]$ mA in steps of 1.22 mA, $V = 0.8$ $\mu$m/s and $b_0 = 100$ $\mu$m. Depending on the value of $I$, the peak(s) of $F$, denoted by $F_\star$, can be positive, negative, or both. In (b), $F_\star$ is plotted as a function of the applied current for different velocities and $b_0 = 100$ $\mu$m.}
		\label{fig:FmaxIs}
	\end{figure*}

	As a last remark about Fig.~\ref{fig:FxL}, it is evident that the inner electrode radius $R_e$ has a minor influence on the adhesion force since there is almost no difference between solid and dashed curves. In addition, taking the limit $R_e \rightarrow 0$ in Eq.~\eqref{forcef} is very useful for obtaining a simplified scenario [Eq.~\eqref{forcef1}] where one can clearly understand the physical contribution of each term to the adhesion force. Nevertheless, as we will analyse $F$ for other values of $V$ and $b_0$ in the following graphs [Figs.~\ref{fig:FmaxDiffVsB0s} and~\ref{fig:FmaxIs}], we will utilize only the full expression~\eqref{forcef}. 
	
	
	The influence of the driving velocity $V$ and initial gap $b_0$ on the adhesion force is investigated in Fig.~\ref{fig:FmaxDiffVsB0s}. In panel (a), we plot $F$ as a function of $L$ for different lifting velocities and two different (constant) values of $I$. Once again, for $I = 15$ mA (solid lines), electro-osmotic and hydraulic forces generate an adhesive force qualitatively similar to the classical case when $I = 0$. As expected, the force is overall larger as the lifting velocity is increased. On the other hand, for $I = -35$ mA~(dashed lines), the scenario can change depending on the driving speed. If the imposed lifting velocity is sufficiently large, e.g., $V =$ 2.4 and 3.2 $\mu$m/s, $F$ is dominated by the hydraulic contribution, and hence is positive. For lower velocities, e.g., $V =$ 0.8 and 1.6  $\mu$m/s, however, $F$ is negative; thus, the electro-osmotic flow overcomes the hydraulic term, and the fluid exerts a repulsive force, which acts to separate the plates spontaneously.

	We move now to Fig.~\ref{fig:FmaxDiffVsB0s}(b), where $F$ is plotted against the nondimensional displacement $L/b_0$ for different values of $b_0$, and two different applied currents, $I = 15$ mA~(solid lines) and $I = -35$ mA~(dashed lines). First, we observe that for all the curves associated with $I = 15$ mA, the force is always positive with its maximum value being a decreasing function of $b_0$. Besides, in contrast to what we have seen in Fig.~\ref{fig:FmaxDiffVsB0s}(a) for different values of $V$, changes in the initial gap $b_0$ can only decrease the amplitude of the overall adhesive force, but never switch its sign. The reason for this behavior is that since $\eta \sigma b^2 \gg 12 \epsilon^2 \zeta^2$, at leading order, $b_0$ contributes equally to the hydraulic and electro-osmotic terms~[see, for simplicity, Eq.~\eqref{forcef1}], so changing its value acts as a prefactor in the force, and, hence, it does not change which term dominates the dynamic.    
	
	Another interesting situation occurs for certain intensities of the electric current in which the applied force $F$ changes its sign during the lifting process. Examples of this behavior is illustrated in figure~\ref{fig:FmaxIs} (a), where $F \times L$ is plotted for $V = 0.80$ $\mu$m, $b_0 = 100$ $\mu$m, and for electric current intensities varying in the range $[-16,-5]$ mA. When $I$ is negative and sufficiently large in magnitude, $F$ is purely negative during the whole lifting process. Nevertheless, as the electric current decreases in magnitude (for instance, in the case of $I = -12.34$ mA), the fluid initially works as an adhesive, with $F>0$ pointing upward. Then, after passing by its maximum value, $F$ decreases and eventually becomes negative. At this moment, instead of acting as an adhesive, the fluid changes its behaviour by exerting a separation force on the plates, and the associated curve displays a negative peak before relaxing asymptotically to zero as $L \rightarrow \infty$. Recall that these changes in $F$ during the lifting occur to keep the upper plate velocity constant.
	
	\begin{figure*}[ht]
		\includegraphics[width=\textwidth]{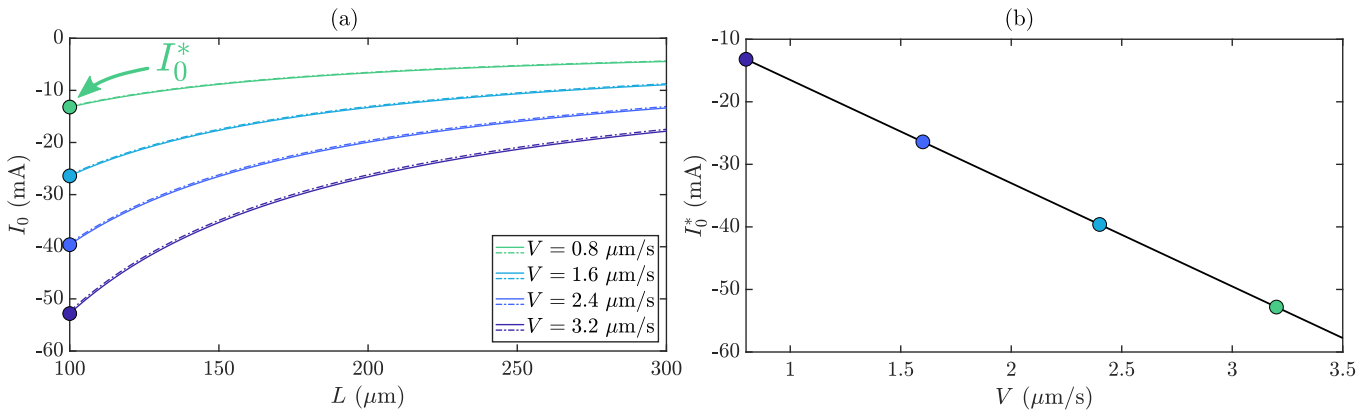}
		\caption{(a) Time-dependent electric current $I_0(t)$ required to vanish the applied force $F$ at all distances $L$, or equivalently at all times. Here, $I_0(t)$ is plotted as a function of $L$ for different lifting velocities and $b_0 = 100$ $\mu$m. Solid lines are obtained from~\eqref{eq:Izero}, while dashed curves are computed with~\eqref{eq:IzeroApprox}. (b) The lowest value of $I_0(t)$ (largest in magnitude), denoted by $I_0^\ast=I_0^{\ast}(V)$, is plotted as a function of the velocity $V$. In panel (a), the arrow points to $I_0^\ast$ for $V=0.8~\mu$m/s.} 
		\label{fig:Izero}
	\end{figure*}

	In Fig.~\ref{fig:FmaxIs}(a), we denote both the positive and negative peaks of the force by $F_{\star}$, which is calculated via $dF/dL|_{L=L_{\star}}=0$, with $F(L_{\star})=F_{\star}$. The behavior of $F_{\star}$ as a function of the electric current for different lifting velocities is shown in Fig.~\ref{fig:FmaxIs}(b). In this panel, first, we observe that the peak of $F$ increases approximately linearly with $I$. Besides, the situation is qualitatively similar for all velocities we used, i.e., for high enough negative currents, the force has a non-zero negative peak, and when the current changes towards positive values, $F$ displays a non-zero positive peak. In addition, between these two ranges of currents, $F$ has a positive and a negative peak, for fixed values of $I$ and $V$. This intermediate range represents the previously mentioned values of $I$, for which the force changes its sign as the plate moves upwards~[see the colourbar in Fig.~\ref{fig:FmaxIs}(a)]. This situation can be seen in the zoomed-in inset in panel (b), which displays the two peaks of $F_\star$: The positive peak arising above a threshold current of $I \approx -13.5$ mA and the negative one vanishing above $I \approx -5$ mA. 
	
	
	It is interesting to note that for a given value of the current, $F_\star$ increases roughly linearly with $V$ when $F_\star > 0$. Nevertheless, for a fixed current, the modulus of $F_\star$ is a decreasing function of the lifting velocity when $F_\star < 0$. This behavior is because $\dot{b}$ only contributes to the hydraulic term of the force. Hence, when the hydraulic contribution prevails in the applied force ($F_\star>0$), one needs to apply a larger force in order to increase the lifting speed. On the other hand, when $I<0$ and the force is dominated by the electric-driven flow ($F_\star<0$), increasing $V$ enhances the counter-balancing hydraulic term, thus decreasing the negative peak of $F$.

	So far, we have verified several situations where one needs to exert a downward force ($F<0$) to keep the lifting velocity constant. This behavior suggests that it is possible to create a dynamical equilibrium between the hydraulic and electro-osmotic contributions, such that the plates separate spontaneously and no applied force is required. This equilibrium can be achieved by imposing a time-dependent electric current $I_0(t)$ such that the electro-osmotic term in Eq.~\eqref{forcef} balances the hydraulic term at all times, i.e., $F(t)=0$ for $I = I_0(t)$. Setting Eq.~\eqref{forcef} to zero, and solving for $I = I_0(t)$, one obtains
	\begin{eqnarray}
		&I_0& (t) = \frac{\dot b}{b^2}\left(\frac{\eta \sigma b^2 - 12\varepsilon^2 \zeta^2}{\varepsilon |\zeta|}\right)\Bigg \{A_e - \frac{b_0^2(A_0 - A_e)^2}{2b^2} \Bigg. \nonumber\\
		&\times& \Bigg.\left[(A_0-A_e)\frac{b_0}{b}- A_e\ln\left(1+\frac{b_0(A_0-A_e)}{A_e b}\right)\right]^{-1} \Bigg \}.\nonumber\\
		\label{eq:Izero}
	\end{eqnarray}
	Note that $F=0$ implies that $L=b$~[see Eq.~\eqref{completeF}], and, thus, we can use $b=b(t) = b_0 + V t$ in Eq.~(\ref{eq:Izero}). Furthermore, by taking the limit $R_e \rightarrow 0$, a considerably simpler expression for $I_0(t)$ is obtained, namely
	\begin{equation}
		I_0 (t) \simeq - \frac{\eta b_0 A_0 V }{2\varepsilon |\zeta|(b_0 + V t)^3}\left[\sigma (b_0 + V t)^2 - \frac{12\varepsilon^2 \zeta^2}{\eta}\right].
		\label{eq:IzeroApprox}
	\end{equation}
	
	In Fig.~\ref{fig:Izero}(a), we depict the evolution of $I_0(t)$ with respect to $L$ for different lifting velocities. In addition, solid (dashed) curves are plotted considering $I_0(t)$ as given by Eq.~\eqref{eq:Izero} [Eq.~\eqref{eq:IzeroApprox}], and we observe that the simpler expression~\eqref{eq:IzeroApprox} is again a very good approximation. To understand the behavior of $I_0(t)$ shown in Fig.~\ref{fig:Izero}(a), first one should note that both the hydraulic and electro-osmotic contributions decay as the distance between the plates increases. However, by inspecting Eq.~(\ref{forcef1}), one verifies that the hydraulic term decreases faster ($\sim b^{-5}$) than the electro-osmotic one ($\sim b^{-4}$). Therefore, during the lifting process, for the electro-osmotic contribution to reduce to the same magnitude as the hydraulic one, the current $I_0(t)$ required to counterbalance the hydraulic term should also diminish as the plates separate. Finally, note that there is an overall decrease in the magnitude of $I_0(t)$ if one wants to lift the upper plate at lower speeds. As a consequence, we verify that even very small electric currents can lift the upper plate when $V$ goes to zero. Nevertheless, such a process would take a considerably long time, becoming useless for any practical purpose.
	
	The largest magnitude of $I_0(t)$ is $|I_0^\ast|= |I_0 (t = 0)|$, which is a linearly decreasing function of the lifting velocity $V$, as shown in Fig.~\ref{fig:Izero}(b). The relevance of $I_0^\ast$ comes in handy when one wishes to lift the upper plate without applying an external force and without imposing a constant velocity. Note that at the very beginning of the lifting process, the upper plate tends to move with velocity $V$ since $I_0^\ast= I_0 (t = 0)$. But because $|I_0^\ast| > |I_0 (t > 0)|$, this velocity is not maintained, and the plate accelerates to higher speeds. Therefore, we verify that $I_0^\ast$ represents the smallest (in magnitude) electric current to detach the plates with a minimum velocity $V$ and without applying an external force.


	\section{Conclusion}
	\label{conclude}
	
	
	
	In this work, we showed that the adhesive strength of a regular liquid sample, such as water and/or glycerol, can be conveniently manipulated by electric means. This electric-tunable adhesive was proposed by taking advantage of the electro-osmotic flow generated by an external electric field along the surfaces that confine the fluid sample in a probe-tack apparatus.
	
	From a modified Darcy's law that considers both hydraulic and electro-osmotic contributions to the fluid flow, we derived the applied force required to pull apart the confining plates at a constant lifting velocity. By controlling the intensity and direction of the applied electric current, we found that the liquid's adhesion force can be enhanced or diminished compared to the traditional, purely viscous case where no electric current (zero electric fields) is applied.

	In addition, when sufficiently large negative electric currents (in the same direction as the lifting-driven flow) are applied, our findings showed that the traditional adhesive behavior of the liquid sample reverses, and the material passes to push apart the probe-tack surfaces rather than holding them together. All these reversibly controllable adhesive features were possible only due to the presence of the electro-osmotic flow, which assists (opposes) the purely viscous force when positive (negative) electric currents are utilized, thus increasing (decreasing) the adhesion force.
	
	We also found that a dynamical equilibrium between viscous and electro-osmotic forces is attained when a specific time-dependent electric current $I_0(t)$ is applied. Under these circumstances, the plates of the probe-tack system separate spontaneously with constant upward velocity without the need to apply a force. Finally, we showed that the largest value, in magnitude, of $I_0(t)$, i.e., $I_0(t=0)=I_0^\ast$, represents the smallest electric current required to separate the plates without an external force and with a minimum upper plate velocity.
	
	Our theoretical work makes specific predictions that have not yet been subjected to experimental checks. Besides its technological relevance, we believe that an eventual experimental verification of our control method would reveal foundational aspects of fluid dynamics related to the interplay between adhesion and electrohydrodynamic phenomena. A natural extension of the current work would be investigating the influence of electric forces on the adhesive properties of complex electric fluids, such as electrorheological fluids suspensions~\citep{ER}, which exhibit an electric field-dependent shear yield stress. Finally, we hope this work will instigate further theoretical and experimental studies on this rich topic.
	
	\begin{acknowledgments}
	E.O.D. acknowledges financial support from Conselho Nacional de Desenvolvimento Cient\'ifico e Tecnol\'ogico (CNPq) through its program 09/2020 (Grant No. 315759/2020-8) and Fundação de Amparo à Ciência e Tecnologia do Estado de Pernambuco (FACEPE) through PPP Project No. APQ-0800-1.05/14.
	\end{acknowledgments}


\begin{thebibliography}{99}
		
		
		
		\bibitem{Gay1} C. Gay and L. Leibler, On stickiness, Phys. Today {\bf 52} (11), 48 (1999).
		
		\bibitem{Aubrey} D. W. Aubrey, {\it Aspects of Adhesion}, edited by K. W. Allen (Elsevier, New York, 1988).
		
		
		\bibitem{Francis}B. A. Francis and R. G. Horn, Apparatus-specific analysis of fluid adhesion measurements, J. Appl. Phys. {\bf 89}, 4167 (2001).
		
		\bibitem{Anke}D. Derks, A. Lindner, C. Creton, and D. Bonn, Cohesive failure of thin layers of soft model adhesives under tension, J. Appl. Phys. {\bf 93}, 1557 (2003).
		
		\bibitem{Gay}S. Poivet, F. Nallet, C. Gay, and P. Fabre, Cavitation-induced force transition in confined viscous liquids under traction, Europhys. Lett. {\bf 62}, 244 (2003).
		
		\bibitem{Dias}E. O. Dias and J. A. Miranda, Variational scheme towards an optimal lifting drive in fluid adhesion, Phys. Rev. E {\bf 86}, 046322 (2012).
		
		\bibitem{Mir7}J. A. Miranda, R. M. Oliveira, and D. P. Jackson, Adhesion phenomena in ferrofluids, Phys. Rev. E {\bf 70}, 036311 (2004).
		
		\bibitem{Mir8}S. A. Lira and J. A. Miranda, Field-controlled adhesion in confined magnetorheological fluids, Phys. Rev. E {\bf 80}, 046313 (2009).
		
		\bibitem{anjosforce} P. H. A. Anjos, E. O. Dias, L. Dias, and J. A. Miranda, Adhesion force in fluids: Effects of fingering, wetting, and viscous normal stresses, Phys. Rev. E {\bf 91}, 013003 (2015).
		
		\bibitem{Russel}M. Tirumkudulu, W. B. Russel, and T. J. Huang, On the measurement of “tack” for adhesives, Phys. Fluids {\bf 15}, 1588 (2003).
		
		\bibitem{Gay2}S. Poivet, F. Nallet, C. Gay, J. Teisseire, and P. Fabre, Force response of a viscous liquid in a probe-tack geometry: Fingering versus cavitation, Eur. Phys. J. E {\bf 15}, 97 (2004).
		
		\bibitem{Mir6}J. A. Miranda, Shear-induced effects in confined non-Newtonian fluids under tension, Phys. Rev. E {\bf 69}, 016311 (2004).
		
		\bibitem{Ben} M. Ben Amar and D. Bonn, Fingering instabilities in adhesive failure, Phys. D {\bf 209}, 1 (2005).
		
		\bibitem{Anke2}A. Lindner, D. Derks, and M. J. Shelley, Stretch flow of thin layers of Newtonian liquids: Fingering patterns and lifting forces, Phys. Fluids {\bf 17}, 072107 (2005).
		
		\bibitem{Anke3}J. Nase, A. Lindner, and C. Creton, Pattern formation during deformation of a confined viscoelastic layer: From a viscous liquid to a soft elastic solid, Phys. Rev. Lett. {\bf 101}, 074503 (2008).
		
		\bibitem{Van}Y. O. M. Abdelhaye, M. Chaouche, and H. Van Damme, The tackiness of smectite muds. 1. The dilute regime, Appl. Clay Sci. {\bf 42}, 163 (2008).
		
		\bibitem{Barral}Q. Barral, G. Ovarlez, X. Chateau, J. Boujlel, B. Rabideau, and P. Coussot, Adhesion of yield stress fluids, Soft Matter {\bf 6}, 1343 (2010).
		
		\bibitem{Randy}R. H. Ewoldt, P. Tourkine, G. H. McKinley, and A. E. Hosoi, Controllable adhesion using field-activated fluids, Phys. Fluids {\bf 23}, 073104 (2011).
		
		
		\bibitem{Diasnew3}E. O. Dias and J. A. Miranda, Effect of fluid inertia on probe-tack adhesion, Phys. Rev. E {\bf 85}, 016312 (2012).
		
		\bibitem{new1}B. Adhikari, T. Howes, B. R. Bhandari, and V. Truong, Stickiness in foods: A review of mechanisms and test methods, Int. J. Food Prop. {\bf 4}, 1 (2001).
		
		\bibitem{new2}Y. O. M. Abdelhaye, M. Chaouche, J. Chapuis, E. Char-laix, J. Hinch, S. Roux, and H. Van Damme, Tackiness and cohesive failure of granular pastes: Mechanistic aspects, Eur. Phys. J. E {\bf 35}, 1 (2012).
		
		\bibitem{new3}O. A. Fadoul and P. Coussot, Saffman-Taylor instability in yield stress fluids: Theory-experiment comparison, Fluids {\bf 4}, 53 (2019).
		
		\bibitem{new4}T. Divoux, A. Shukla, B. Marsit, Y. Kaloga, and I. Bischof-berger, Criterion for fingering instabilities in colloidal gels, Phys. Rev. Lett. {\bf 124}, 248006 (2020).
		
		\bibitem{new5}X. Zhang, O. Fadoul, E. Lorenceau, P. Coussot, Yielding and flow of soft-jammed systems in elongation, Phys.Rev. Lett. {\bf 120}, 048001 (2018).
		
	    \bibitem{new6}M. J. Hayes and M. I. Smith, Slip in adhesion tests of a Kaolin clay, Eur. Phys. J. E {\bf 44}, 1 (2021).
		
		\bibitem{Zosel} A. Zosel, Adhesion and tack of polymers: Influence of mechanical properties and surface tensions, Colloid Polym. Sci. {\bf 263}, 541 (1985).
		
		\bibitem{Lakrout} H. Lakrout, P. Sergot, and C. Creton, Direct observation of cavitation and fibrillation in a probe tack experiment on model acrylic pressure-sensitive-adhesives, J. Adhes. {\bf 69}, 307 (1999).
		
		\bibitem{Bazant1} M. Mirzadeh and M. Z. Bazant, Electrokinetic control of viscous fingering, Phys. Rev. Lett. {\bf 119}, 174501 (2017).
		
		\bibitem{Bazant2} T. Gao, M. Mirzadeh, P. Bai, K. M. Conforti, and M. Z. Bazant, Active control of viscous fingering using electric fields, Nat. Commun. {\bf 10}, 4002 (2019).
		
		\bibitem{Bazant3}M. Mirzadeh, T. Zhou, M. A. Amooie, D. Fraggedakis, T. R. Ferguson, and M. Z. Bazant, Vortices of electro-osmotic flow in heterogeneous porous media, Phys. Rev. Fluids {\bf 5}, 103701 (2020).
		
		\bibitem{pedroeletro}P. H. A. Anjos, M. Zhao, J. Lowengrub, and S. Li, Electrically controlled self-similar evolution of viscous fingering patterns, Phys. Rev. Fluids {\bf 7}, 053903 (2022).
		
		\bibitem{mengeletro}M. Zhao, P. H. A. Anjos, J. Lowengrub, W. Ying, and S. Li, Numerical study on viscous fingering using electric fields in a Hele-Shaw cell, https://arxiv.org/abs/2201.05956 (2022).
		
		\bibitem{EDL1}B. J. Kirby and E. F. Hasselbrink, Zeta potential of microfluidic substrates: 1. Theory, experimental techniques, and effects on separations, Electrophoresis {\bf 25}, 187 (2004).
		
		\bibitem{EDL2}R. J. Hunter, {\it Zeta Potential in Colloid Science: Principles and Applications}, 3rd ed. (Academic Press, New York, 1988).
		
		\bibitem{EDL3}R. J. Hunter, {\it Foundations of Colloid Science}. (Oxford University Press, New York, 2001).
		
	
		
		\bibitem{Leider}P. J. Leider, Squeezing flow between parallel disks. II. Experimental results, Ind. Eng. Chem. Fundam. {\bf 13}, 342 (1974).
		
		\bibitem{Anke4}A. Lindner, D. Bonn, and J. Meunier, Viscous fingering in a shear-thinning fluid, Phys. Fluids {\bf 12}, 265 (2000).
		
		\bibitem{ER}T. Hao, Electrorheological fluids, Adv. Mater. {\bf 13(24)}, 1847 (2001).
	
	\end{thebibliography}
\end{document}